\documentclass[11pt]{article}
\usepackage{geometry}                % See geometry.pdf to learn the layout options. There are lots.
\usepackage{graphicx}
\usepackage{amsmath}
\usepackage{amssymb}
\title{Topology in Physics\footnote{TOP 2005 Symposium, Sapporo, Japan, March 2005.}}
\author{R. Jackiw\\
\it \small Center for Theoretical Physics\\
\it \small MIT, Cambridge, MA 02139-4307}
\date{\it\small MIT-CTP-3608}                                           % Activate to display a given date or no date
\begin{document}
\maketitle
\begin{abstract}
 The phenomenon of quantum number fractionalization is explained. The relevance of non-trivial phonon field topology is emphasized.
\end{abstract}
%\rightline{\tiny \it MIT-CTP-3608}
\section{Introduction}
%\subsection{}
Discussions of the spatial forms of physical materials use in a natural way geometrical and topological concepts. It is to be expected that arrangements of matter should form patterns that are described by pre-existing mathematical structures drawn from geometry and topology. But theoretical physicists also deal with abstract entities, which do not have an  actual material presence. Still geometrical and topological considerations  are relevant to these ephemeral theoretical constructs. I have in mind fields, both classical and quantum, which enter into our theories of fundamental processes. These fields $\phi (x)$ provide a mapping from a ``base" space or space-time on which they are defined into the field ``target" manifold on which they range. The base and target spaces, as well as the mapping, may possess some non-trivial topological features, which affect the fixed time description and the temporal evolution of the fields, thereby influencing the physical reality that these fields describe. Quantum fields of a quantum field theory are operator valued distributions whose relevant topological properties are obscure. Nevertheless, topological features of the corresponding classical fields are important in the quantum theory for a variety of reasons: (i) Quantized fields can undergo local (space-time dependent) transformations (gauge transformations, coordinate diffeomorphisms) that involve classical functions whose topological properties determine the allowed quantum field theoretic structures. (ii) One formulation of quantum field theory uses a functional integral over classical fields, and classical topological features become relevant. (iii) Semi-classical (WKB) approximations to the quantum theory rely on classical dynamics, and again classical topology plays a role in the analysis.

Topological effects in quantum electrodynamics were first appreciated by Dirac in his study of the  quantum mechanics for (hypothetical) magnetic monopoles. This analysis leads directly to  contemporary analysis of Yang-Mill theory -- the contemporary generalization of Maxwell's electrodynamics -- and has yielded several significant results: the discovery of the $\theta$-vacuum angle; the recognition that c-number parameters in the theory may require quantization for topological reasons (like Dirac's monopole strength); the realization that the chiral anomaly equation is just the local version of the celebrated Atiyah-Singer index theorem.

Here I shall not describe the Yang-Mills investigations;  they are too technical and too specialized for this general audience. Rather I shall show you  how a topological effect in a condensed matter situation leads to charge fractionalization. This phenomenon has a physical realization in 1-dimensional (lineal) polymers, like polyacetylene, and in 2-dimensional (planar) systems, like the Hall effect.

The polyacetylene story is especially appealing, because it can be told in several ways:  in pictorial terms which only involves counting, or in the first quantized formalism for quantum mechanical equations, or in the second quantized formalism of a quantum field theory \cite{rj1}.

\section{The Polyacetylene Story (Counting Argument)}
Polyacetylene is a material consisting of parallel chains of carbon atoms, with electrons moving primarily along the chains, while hopping between chains is strongly suppressed. Consequently, the system is effectively 1-dimensional.  The distance between carbon atoms is about 1\AA.

If  the atoms are considered to be completely stationary, {\it i.e.} rigidly attached to their equilibrium lattice sites, electron hopping along the chain  is a structureless phenomenon.
\begin{figure}[htbp] %  figure placement: here, top, bottom, or page
   \centering
   \includegraphics[scale=.26]{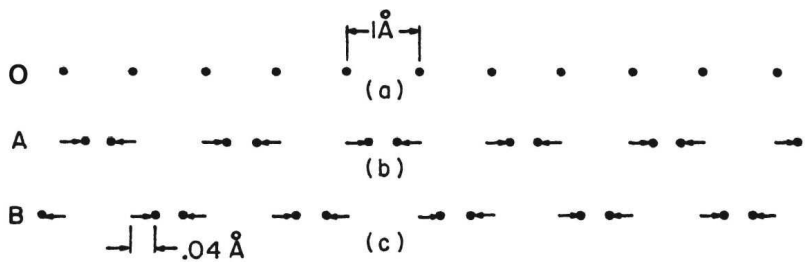} 
   \caption{(a) The rigid lattice of polyacetylene; (O) the carbon atoms are equally spaced 1 \AA\, apart. (b), (c) The effect of Peierls' instability is to shift the carbon atoms .04\AA to the right (A) or to the left (B), thus giving rise to a double degeneracy.}
   \label{fig:example1}
\end{figure}
However, the atoms can oscillate around their rigid lattice positions for a variety of reasons, like zero-point motion, thermal excitation, etc. It might be thought that these effects merely give rise to a slight fuzzing of the undistorted-lattice situation.

In fact this is not correct; something more dramatic takes place. Rather than oscillating about the rigid-lattice sites, the atoms first shift a distance of about .04 \AA\ and then proceed to oscillate around the new, slightly distorted location. That this should happen was predicted by Peierls, and is called the Peierls instability.  Due to reflection symmetry, there is no difference between a shift to the right or a shift to the left; the material chooses one or the other, thus breaking spontaneously the reflection symmetry, and giving rise to doubly degenerate vacua, called A and B. 
%One represents chemical bonding pattern by a double bond connecting atoms that are closer together, and the single bond connecting those that are further apart.

If the displacement is described by a field $\phi$ which depends on the position x along the lattice, the so-called phonon field, then Peierls' instability, as well as detailed dynamical calculations indicate that the energy density $V(\phi)$, as a function of constant $\phi$, has a double-well shape. 
\begin{figure}[htbp] %  figure placement: here, top, bottom, or page
   \centering
   \includegraphics[scale=.16]{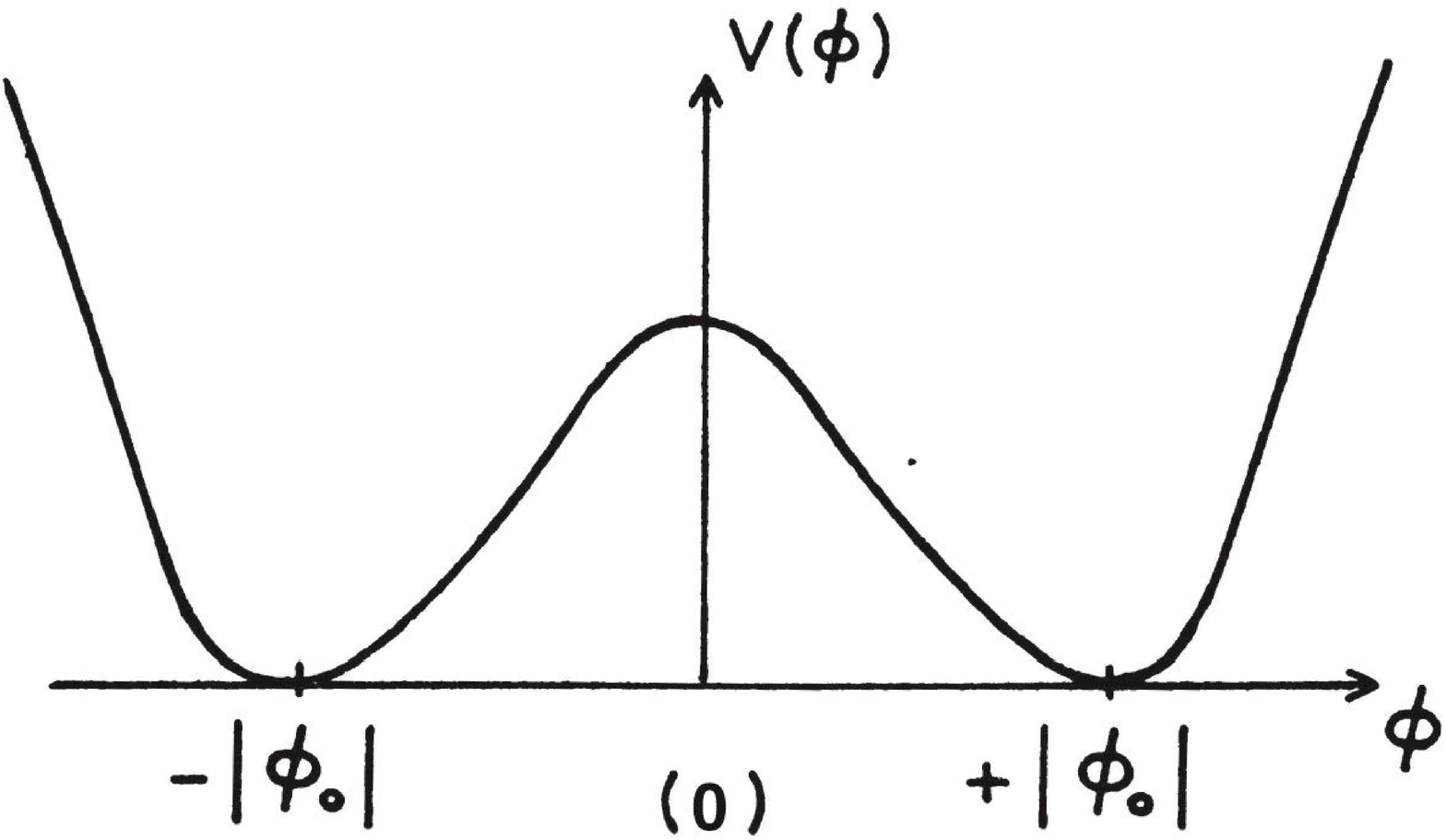} 
   \caption{Energy density $V(\phi)$, as a function of a constant phonon field $\phi$. The symmetric stationary point, $\phi = 0$, is unstable. Stable vacua are at $\phi = {\scriptstyle{+}} |\phi_0|, \text{(A) and}\,  \phi = \text{-}|\phi_0|, \text{B}$. }
   \label{fig:example2}
    \includegraphics[scale=.22]{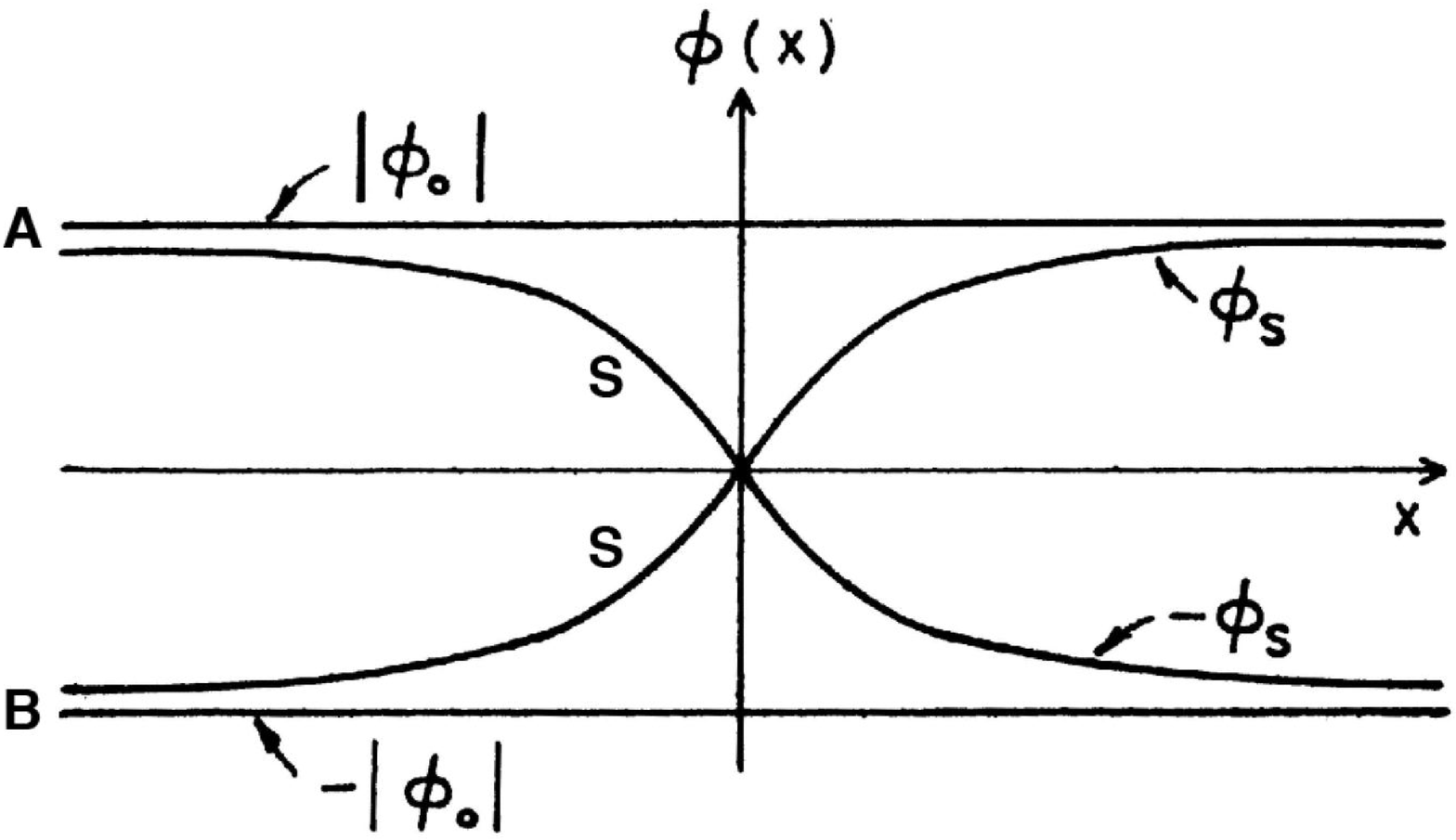} 
\caption{The two constant fields, $\pm \mid\phi_0 \mid$, correspond to the two vacua (A and B). The two kink fields, $\pm \phi_s$, interpolate between the vacua and represent domain walls.}
   \label{fig:En}
\end{figure}
The symmetric point $\phi=0$ is unstable; the system in its ground state must choose one of the two equivalent ground states $\phi= \pm \mid \phi_0 \mid = \pm .04$\AA. In the ground states, the phonon field has uniform values, independent of x.

By now it is widely appreciated that whenever the ground state is degenerate there frequently exist additional stable states of the system, for which the phonon field is non-constant. Rather, as a function of x, it interpolates, when x passes from negative to positive infinity, between the allowed ground states. These are the famous solitons, or kinks. For polyacetylene they correspond to domain walls which separate regions with vacuum A from those with vacuum B, and vice versa. One represents the chemical bonding pattern by a double bond connecting atoms that are closer together, and the single bond connecting those that are further apart.
\begin{figure}[htbp] %  figure placement: here, top, bottom, or page
   \centering
   \includegraphics[scale=.38]{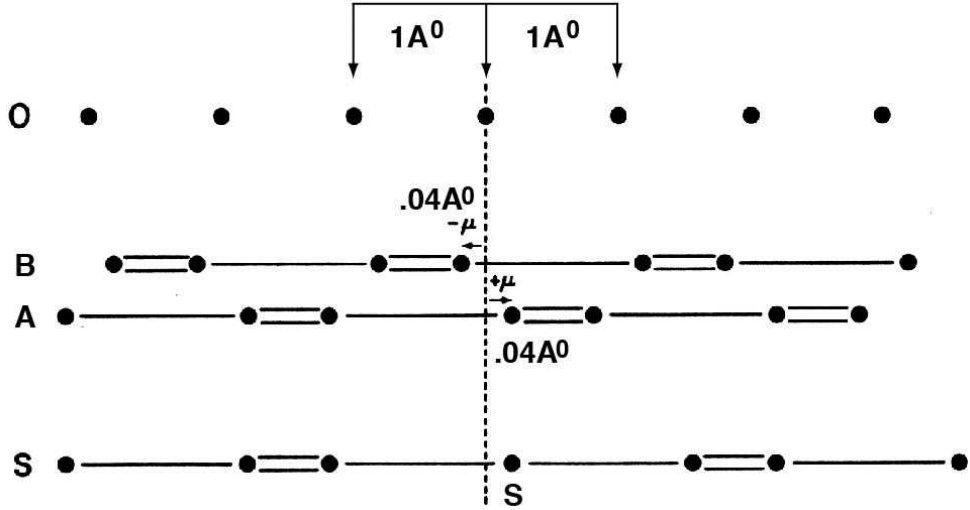} 
   \caption{Polyacetylene states. The equally spaced configuration (O) possesses a left-right symmetry, which however is energetically unstable. Rather in the ground states the carbon atoms shift a distance $\mu$ to the left or right, breaking the symmetry and producing two degenerate vacua (A, B). A soliton (S) is a defect in the alteration pattern; it provides a domain wall between configurations (A) and (B). 
   %a microscopic Hamiltonian for the system has been proposed by Su, Schrieffer, and Heeger (SSH). In the continuum limit, electron transport is governed by a Dirac-type Hamiltonian in one dimension
   }
   \label{fig:example4}
\end{figure}

Consider now a polyacetylene sample in the A vacuum, but with two solitons along the chain. Let us count the number of links in the sample without solitons and compare with number of links where two solitons are present. It suffices to examine the two chains only in the region where they differ, {\it i.e.} between the two solitons. Vacuum A exhibits 5 links, while  the addition of two  solitons decreases the number of links to 4. The two soliton state exhibits a deficit of one link. If  now we imagine separating the two solitons a great distance, so that they act independently of one another, then each soliton carries a deficit of half a link, and the quantum numbers of the link, for example the charge, are split between the two states.  This is the essence of fermion fractionization.
\begin{figure}[htbp] %  figure placement: here, top, bottom, or page
   \centering
   \includegraphics[scale=.25]{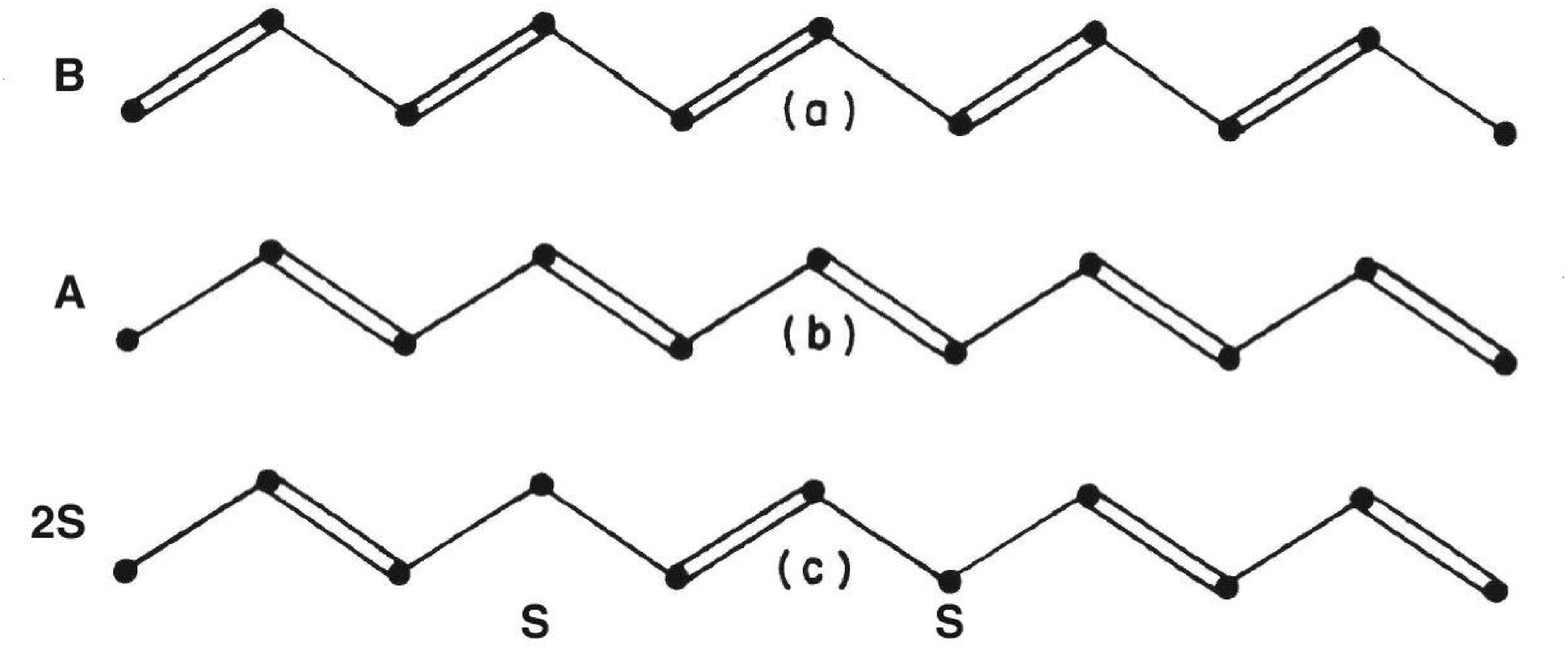} 
   \caption{(a), (b) Pattern of chemical bonds in vacua A and B. (c) Two solitons inserted into vacuum A.}
   \label{fig:example5}
\end{figure}

It should be emphasized that we are not here describing the familiar situation of an electron moving around a two-center molecule, spending ``half" the time with one nucleus  and ``half" with the other. Then one might say that the electron is split in half, on the average;  however fluctuations in any quantity are large. But in our soliton example, the fractionization is without fluctuations; in the limit of infinite separation one achieves an eigenstate with fractional eigenvalues. 

We must however remember that the link in fact corresponds to two states: an electron with spin up and another with spin down. This doubling obscures the dramatic charge $\frac{1}{2}$ effect, since everything  must be multiplied by 2 to account for the two states. So in polyacetylene, a soliton carries a charge deficit  of one unit of electric charge. Nevertheless charge fractionization leaves a spur: the soliton state has net charge, but no net spin, since all of the electron spins are paired. If  an additional electron is inserted into the sample, the charge deficit is extinguished, and one obtains a neutral state, but now there is a net spin. These spin-charge assignments (charged -- without spin, neutral -- with spin) are unexpected, but in fact have been observed, and provide experimental verification for the soliton picture and fractionalization in polyacetylene.

Notice that in this simple counting argument no mention is made of topology. This feature emerges only when an analytic treatment is given. I now turn to this.

\section{The Polyacetylene Story (Quantum Mechanics)}
I shall now provide a calculation which shows how charge $1/2$ arises in the quantum mechanics of fermions in interaction with solitons. The fermion dynamics are governed by an one-dimensional Dirac Hamiltonian, $H(\phi)$, which also depends on a background phonon field $\phi$, with which the fermions intact. The Dirac Hamiltonian arises not because the electrons are relativistic. Rather it emerges in a certain well-formulated approximation to the microscopic theory,  which yields a  quantal equation that is a 2x2 matrix equation, like a Dirac equation. In the vacuum sector, $\phi$ takes on a constant value $\phi_0$, appropriate  to the vacuum. When a soliton is present, $\phi$ becomes the appropriate, static soliton profile $\phi_s$. We need not be any more specific. We need not insist on any explicit soliton profile. All that we require is that the topology [{\it i.e.} the large distance behavior] of the soliton profile be non-trivial. 

In the present lineal case the relevant topology is that infinity corresponds to two points, the end points of the line, and the phonon field in the soliton sector behaves differently at the points at infinity.

To analyze the system we need the eigenmodes, both in the vacuum and soliton sectors.
\begin{eqnarray}
H(\phi_0) \psi^v_E &=& E \psi^v_E \label{fig1}\\
H(\phi_s) \psi^s_E &=& E \psi^s_E
\label{fig2}
\end{eqnarray}
The Dirac equation is like a matrix-valued ``square root" of the wave equation. Because a square root is involved, there will be in general negative energy solutions and positive energy solutions. The negative energy solutions correspond to the states in the valence band; the positive energy ones, to the conduction band. In the ground state, all the negative energy levels are filled, and the ground state charge is the integral over all space of the charge density $\rho (x)$, which in turn is constructed from all the negative energy wave functions.
%\begin{eqnarray}
%\rho(x) = \int^0\limits_{- \infty} d E \, \rho_{E} \, (x) \nonumber \\[.50ex]
%\rho_E (x) = \psi^\ast_E \, (x)\, \psi_E\, (x) \label{fig3}
%\end{eqnarray}
\begin{equation}
\rho(x) = \int^0\limits_{- \infty} d E \, \rho_{E} \, (x), \  
\rho_E (x) = \psi^\ast_E \, (x)\, \psi_E\, (x) \label{fig3}
\end{equation}
Of course integrating (\ref{fig3}) over $x$  will produce an infinity; to renormalize we measure all charges relative to the ground state in the vacuum sector. Thus the soliton charge is
\begin{equation}
Q = \int \, dx \, \int^0\limits_{- \infty} dE\, \{\rho^s_E\, (x) - \rho^v_E\, (x)\}
\label{fig4}
\end{equation}
Eq. (\ref{fig4}) may be completely evaluated without explicitly specifying the soliton profile, nor actually solving for the negative energy modes, provided H possesses a further property. We assume that there exists a conjugation symmetry which takes positive energy solutions of (\ref{fig1}) and (\ref{fig2}) into negative energy solutions. (This is true for polyacetylene.) That is, we assume that there exists a unitary 2x2 matrix $M$, such that
\begin{equation}
M \psi_E = \psi_{- E}
\label{fig5}
\end{equation}
An immediate consequence, crucial to the rest of the argument, is that the charge density at $E$ is an even function of $E$.
\begin{equation}
\rho_{\scriptscriptstyle E}  (x) = \rho_{\scriptscriptstyle - E}(x)
\label{fig6}
\end{equation}

Whenever one solves a conjugation symmetric Dirac equation, with a topologically interesting background field, like a soliton, there always are, in addition to the positive and negative energy solutions related to each other by conjugation, self-conjugate, normalizable zero-energy solutions. That this is indeed true can be seen by explicit calculation. However, the occurrence of the zero mode is also predicted by very general mathematical theorems about differential equations. These so-called ``index theorems" count the zero eigenvalues, and insure that the number is non-vanishing whenever the topology of the background is non-trivial. We shall assume that there is just one zero mode, described by the normalized wave function $\psi_0$. 

To evaluate the charge $Q$ in (\ref{fig4}), we first recall that the wave functions are complete, both in the soliton sector and in the vacuum sector.
\begin{equation}
\int^\infty\limits_{- \infty} d E\, \psi^\ast_E \, (x)\, \psi_E (y) = \delta (x-y)
\label{fig7}
\end{equation}
As a consequence, it follows that
\begin{subequations}
\begin{equation}
\int^\infty\limits_{- \infty} d E\, [\rho^s_E \, (x) \, -\rho^v_E\, (x)] = 0
\label{fig8}
\end{equation}
In the above completeness integral over all energies, we record separately the negative energy contributions, the positive energy contributions, and for the soliton, the zero-energy contribution. Since the positive energy charge density is equal to the negative one, by virtue of (\ref{fig6}), we conclude that (\ref{fig8}) may be equivalently written as an integral over negative $E$.
\begin{equation}
\int^0\limits_{- \infty} d E\, [2\rho^s_E\, (x)  - 2\rho^v_E\, (x)] + \psi^\ast_0\, (x) \, \psi_0 \, (x) =0 
\end{equation}
\end{subequations}
Rearranging terms give
\begin{equation}
Q= \int d x \int^0\limits_{-\infty} d E [\rho^s_E (x) - \rho^v_0 (x)] = -\frac{1}{2} \int d x \psi_0 (x) \psi_0 (x) = -\frac{1}{2}
\label{fig9}
\end{equation}

This is the final result: the soliton's charge is $-\frac{1}{2}$; a fact that follows from completeness (\ref{fig7}) and conjugation symmetry (\ref{fig6}). It is seen in  (\ref{fig9}) that the zero-energy mode is essential to the conclusion. The existence of the zero mode in the conjugation symmetric case is assured by the non-trivial topology of the background field. The result is  otherwise completely general.

\section{The Polyacetylene Story (Quantum Field Theory)}
The quantum mechanical derivation that I just presented does not address the question of whether the fractional half-integer charge is merely an uninteresting expectation value or whether it is an eigenvalue. To settle this, we need a quantum field theory approach, that is we need to second quantize the field. For this, we expand $\Psi$, which now is an anti-commuting quantum field operator, in eigenmodes of our Dirac equation in the soliton sector as 
\begin{eqnarray}
\Psi &=& \sum (b_E \, \psi^s_E + d^\dagger_E \, \psi^s_{-E}) + a \psi_0 \nonumber\\
\Psi^\dagger&=& \sum\limits^E (b^\dagger_E \, \psi^{s \ast}_E + d_E \, \psi^{s \ast}_{-E}) + a^\dagger \psi_0
\label{fig10}
\end{eqnarray}
The important point is that while the finite energy modes $\psi^s_{\pm E}$ enter with annihilation particle (conduction band) operators $b_E$ and creation anti-particle (valence band) operators $d^\dagger_E$, the zero mode does not have a partner and is present in the sum simply with the operator $a$. The zero energy state is therefore doubly degenerate. It can be empty $\mid->$, or filled $\mid+>$, and the $a, a^\dagger$ operators are realized as
%\begin{eqnarray}
%a \mid+> &=& \mid - > \nonumber\\
%a^\dagger \mid+> &=& 0\nonumber\\
%a \mid-> &=& 0 \nonumber\\
%a^\dagger \mid+> &=& \mid + >
%\label{fig11}
%\end{eqnarray}
\begin{equation}
a \mid+> = \mid - >,\ a^\dagger \mid+> = 0,\ a \mid-> = 0,\ a^\dagger \mid+> = \mid + >
\label{eq11}
\end{equation}

The charge operator $Q = \int d x \psi^\dagger \psi$ must be properly defined to avoid infinities. This is done, according to Schwinger's prescription in the vacuum sector, by replacing the formal expression by 
\begin{equation}
Q = \frac{1}{2} \int d x \, (\psi^\dagger \psi - \psi \psi^\dagger)
\label{eq12}
\end{equation}
We adopt the same regularization prescription for the soliton sector and insert our expansion (\ref{fig10}) into (\ref{eq12}). We find with the help of the orthonormality of wave functions
\begin{eqnarray}
Q &=& \frac{1}{2} \, \sum\limits_E \, (b^\dagger_E \, b_E + d_E\, d^\dagger_E - b_E \, b^\dagger_E - d^\dagger_E \, d_E)  + \frac{1}{2} (a^\dagger a - a a^\dagger)\nonumber \\
&=& \sum\limits_E (b^\dagger_E\, b_E - d^\dagger_E \, d_E) + a^\dagger a - \frac{1}{2} 
\label{eq13}
\end{eqnarray}
Therefore the eigenvalues for $Q$ are
\begin{equation}
Q \mid-> = -\frac{1}{2} |->, \ \ Q \mid+> = \frac{1}{2}  |+> \text{!}
\label{eq14}
\end{equation}

\section{Conclusion}
This then concludes my polyacetylene story, which has experimental realization and confirmation. And the remarkable effect arises from the non-trivial topology of the phonon field in the soliton sector.

Many other topological effects have been found in the field theoretic descriptions of condensed matter and particle physics.  Yet we must notice that mostly these arise in phenomenological descriptions, not in the fundamental theory. In condensed matter the fundamental equation is the many-body Schr\"{o}dinger equation with Coulomb interactions. This does not show any interesting topological structure. Only when it is replaced by effective, phenomenological equations do topological considerations become relevant for the effective description. Fundamental (condensed matter) Nature is simple! 

Similarly in particle physics, our phenomenological, effective theories, like the Skyrme model, enjoy a rich topological structure. Moreover, even  the Yang-Mills theory of our fundamental  ``standard particle physics model"  supports non-trivial topological structure, which leads to the QCD vacuum angle. In view of my previous observation, can we take this as indirect evidence that thisYang-Mills based theory also is a phenomenological, effective description and at a more fundamental level -- yet to be discovered -- we shall find a simpler description that does not have any elaborate mathematical structure. Perhaps in this final theory Nature will be described by simple counting rules -- like my first polyacetylene story. Surely this will not be the behemoth of string theory.

This work is supported in part by funds provided by the U.S. Department of Energy (D.O.E.) under cooperative research agreement DE-FC02-94ER40818.

%\section*{Conclusion}
%Topology is relevant only to effective, phenomenological descriptions  in condensed matter, whose fundamental equation is the many body Schwinger equation with Contomb interaction. In particle physics phenomenological theories, like the Skyrme model, enjoy a rich topological structure. But also our fundamental Yang-Mills theory has topological effects (QCD vacuum angle).

%\uppercase{Question}:  Is Yang-Mills theory phenomenological?

%\newpage
%.

\begin{thebibliography}{99}
\bibitem{rj1}
This research was performed in collaboration with C. Rebbi, and independently by W.P. Su, J.R. Schrieffer and A. Heeger. For a summary see R. Jackiw and  J.R. Schrieffer ``Solitons with Fermion number {\small 1/2} in Condensed Matter and Relativistic Field Theories" {\it Nucl. Phys.} {\bf B190}, 253 (1981).
\end{thebibliography}
\end{document}